\documentclass{article}
\pdfoutput=1
\usepackage[utf8]{inputenc}
\usepackage{amsfonts}
\usepackage{mathtools}
\usepackage{url}

\title{Definition of geometric space around analytic fractal trees using derivative coordinate functions}
\author{
        Henk Mulder \\
        henk.mulder@geneticfractals
}
\date{\today}
\date{March 2017}

\usepackage{graphicx}
\usepackage{amssymb}

\newtheorem{theorem}{Theorem}[section]

\newtheorem{definition}[theorem]{definition}

\begin{document}

\maketitle

\begin{abstract}
The concept of derivative coordinate functions proved useful in the formulation of analytic fractal functions to represent smooth symmetric binary fractal trees [1]. In this paper we introduce a new geometry that defines the fractal space around these fractal trees. We present the canonical and degenerate form of this fractal space and extend the fractal geometrical space to $\mathbb{R}^3$ explicitly and $\mathbb{R}^n$ by a recurrence relation. We also discuss the usage of such fractal geometry .
\end{abstract}

\section{Introduction}

Derivative coordinate functions allow us to define multivalued analytic functions to represent smooth binary symmetric tree fractals as sets of partially overlapping analytical paths [1]. We found that with this formulation we can express a binary symmetric tree fractals as
\begin{equation}
\check{p}(s)=\int \frac{\partial r}{\partial s} e^{i \check{u}(s) \int \frac{\partial \phi}{\partial s} ds} ds
\end{equation}
Where $\frac{\partial r}{\partial s}$ and $\frac{\partial \phi}{\partial s}$ are the derivative coordinate functions and $\check{u}(s)$ is the multivalued unit function that has values $\{-1,1\}$ except at branch points when $\check{u}(s)=\{0\}$. Under integration this generates left and right branches starting from each branch point [1].

We note that the check mark over a variable indicates that this is a fractal variable which represents a set of values that correspond to a specific position along a set of branches of the same generation. For example, the midpoint between the 3rd and 4th generation of branches is a set of $2^{3}$ values. We also state that such fractal variable is by definition dependent on the path variable $s$ and will not always specifically state the path variable. For example $\check{u}\equiv\check{u}(s)$.

To simplify the notation we use the dot notation which is justified if we recognize that $\dot{r}(s)$ is a velocity function along the fractal paths and $\dot{\varphi}$ is the angular frequency, i.e. relative change of angle. So equation 1 is written as
\begin{equation}
\check{p}(s)=\int \dot{r} e^{i \check{u} \int \dot{\varphi} ds} ds
\end{equation}

To remind ourselves what this represents, below is an example of a tree fractal that uses derivative coordinate functions $\dot{\varphi}=\frac{\pi}{3}$ and  $\dot{r}=(\frac{2}{3})^s$. 
\begin{figure}[!ht]
  \includegraphics[width=1.0\textwidth]{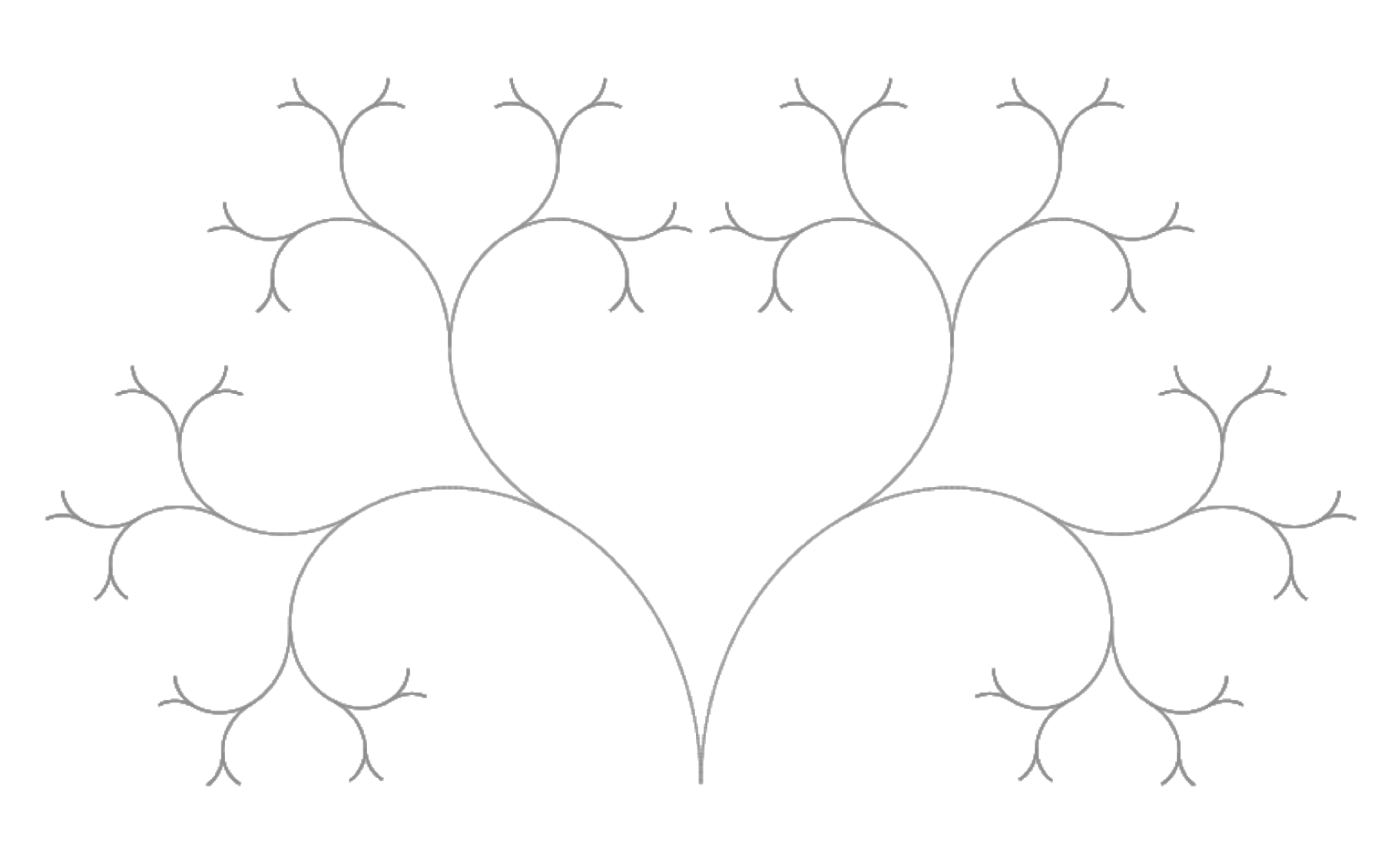}
  \caption{smooth tree fractal using derivative coordinate functions $\dot{r}(s)$ and $\dot{\varphi}(s)$}
\end{figure}

This tree fractal was generated with the smooth tree fractal explorer [2].

\section{Introducing a new geometry}
Using derivative coordinate functions and the formulation of analytic tree fractals, we can now develop a new geometric that is centered around the fractal paths, i.e. the branches. To facilitate the discussion, we restrict ourselves to a single branch of a smooth tree fractal and then extend this to the full tree fractal.

The general definition of smooth tree fractals generates arbitrary binary symmetric tree fractals but we will limit ourselves to the most basic smooth tree fractal, such as shown in Figure 1. This is given by

\begin{equation}
\check{p}(s)=\int R^{s} e^{i \check{u}\Phi s} ds
\end{equation}

where $R$ and $\Phi$ are constant. This represents a smooth tree fractal with branches that shorten (or lengthen) exponentially and where the branches curl constantly. 

When considering a single branch, we can drop the unit branch function $\check{u}(s)$ and for $\Phi<0$ get the path function of the most right hand branch

\begin{equation}
p(s)=\int R^{s} e^{i\Phi s} ds
\end{equation}

We define the new geometry by introducing a coordinate pair $(x_{1},x_{2})$ with the condition that $x_{2}$ is perpendicular to $x_{1}$, i.e. its path of progression is rotated by $\pi/2$ with respect to the path of progression of $x_{1}$. To project a point $x=(x_{1},x_{2})$ on the fractal space, we first move $p$ in the $x_{1}$ direction and then in the $x_{2}$ direction. We can represent this as a path integral. In general we can define a point found by coordinates along a path function as
\begin{definition}
A point $f(x)$ determined by coordinates $x=(x_{1},x_{2}) \in \mathbb{R}^{2}$ such that $x_{1}$ progresses along a path $p(s)$ from $p(0)$ to $p(x_{1})$ and then rotated by $\pi/2$, continuous by $x_{2}$ along a path $p(s)$ from $p(x_{1})$ to $p(x_{1}+x_{2})$ is given as
\begin{equation}
f(x)=\int_{0}^{x_{1}}p(s)ds + i\int_{x_{1}}^{x_{2}+x_{1}}p(s)ds
\end{equation}
\end{definition}

\begin{figure}[!ht]
\centering
  \includegraphics[width=0.8\textwidth]{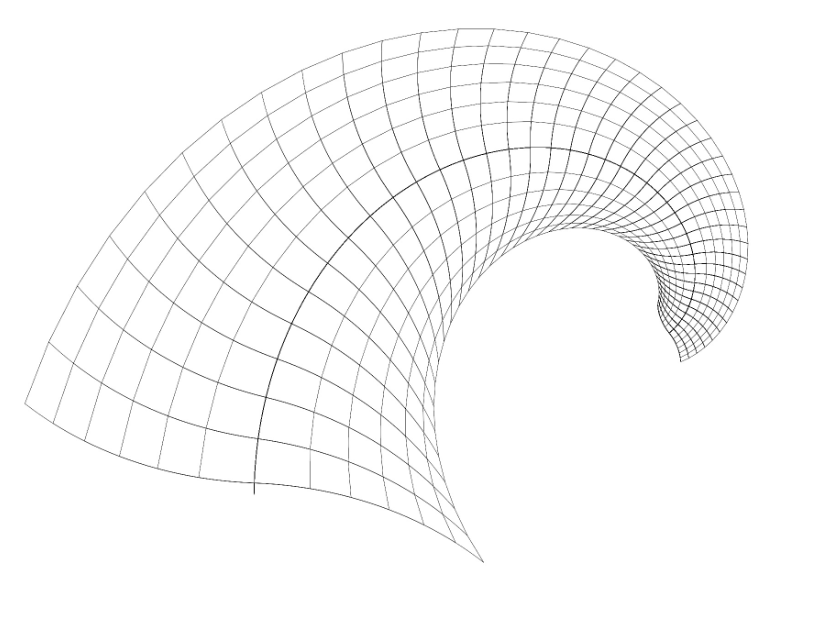}
  \caption{geometrical grid along one of the fractal branches}
\end{figure}

For the fractal branch under consideration, we can substitute $p(s)$,

\begin{equation}
f(x)=\int_{0}^{x_{1}}R^{s}e^{i\Phi s}ds + i\int_{x_{1}}^{x_{1}+x_{2}}R^{s}e^{i\Phi s}ds
\end{equation}

With (6) we can now calculate any point $x$ with respect to the fractal. Figure 2 shows a coordinate grid for whole values of $x_{1}$ and $x_{2}$. We have chosen the most right hand branch of the tree fractal.

We can now extend this new geometry to all branches of the fractal by reintroducing the fractal function $\check{p}(s)$and define the fractal space:

\begin{definition}
The fractal geometric space $\check{F}(x)$ centered around a fractal $\check{p}(x)$ for all coordinates $x=(x_{1},x_{2}) \in \mathbb{R}^{2}$ is defined as
\begin{equation}
\check{F}(x)=\int_{0}^{x_{1}}\check{p}(s)ds + i\int_{x_{1}}^{x_{2}+x_{1}}\check{p}(s)ds
\end{equation}
\end{definition}

Using fractal functions with derivative coordinate functions, we obtain the fractal geometric space
\begin{equation}
\check{F}(x)=\int_{0}^{x_{1}}\dot{r}(s)e^{i\dot{\varphi}(s)\check{u} s}ds + i\int_{x_{1}}^{x_{2}+x_{1}}\dot{r}(s)e^{i\dot{\varphi}(s)\check{u} s}ds
\end{equation}

We are reminded that these integrals return values for each branch path as determined by specific branch instances of the unit fractal function. For example when $u(s)={-1,-1,-1,-1,...}$, the right hand spiralling branch is generated. Note the missing check mark on $u(s)$ indicating that this is a path function, not a fractal function. When $u(s)=\{-1,1,-1,1,...\}$, the central right "snaking" path is generated.

Figure 3 shows an example of a the coordinate grid for the complete tree fractal.
\section{Function plots in fractal geometric space} 
Using the definition of the fractal geometric space, we can plot images of well known objects such as conics. Figure 3 shows an ellipse plotted on the new geometry for the first 4 generations of all branches of an analytic tree fractal. We can see the key features of the new geometry: 
\begin{itemize}
  \item Shrinkage of the scale as the ellipse progresses along the fractal.
  \item Ever sharper bending of the fractal branches 
  point.
  \item Symmetric replication of the ellipse at each branch point, starting from a single point of at the bottom edge of an ellipse at the trunk of the tree fractal.
\end{itemize}

\begin{figure}[!ht]
  \includegraphics[width=1.0\textwidth]{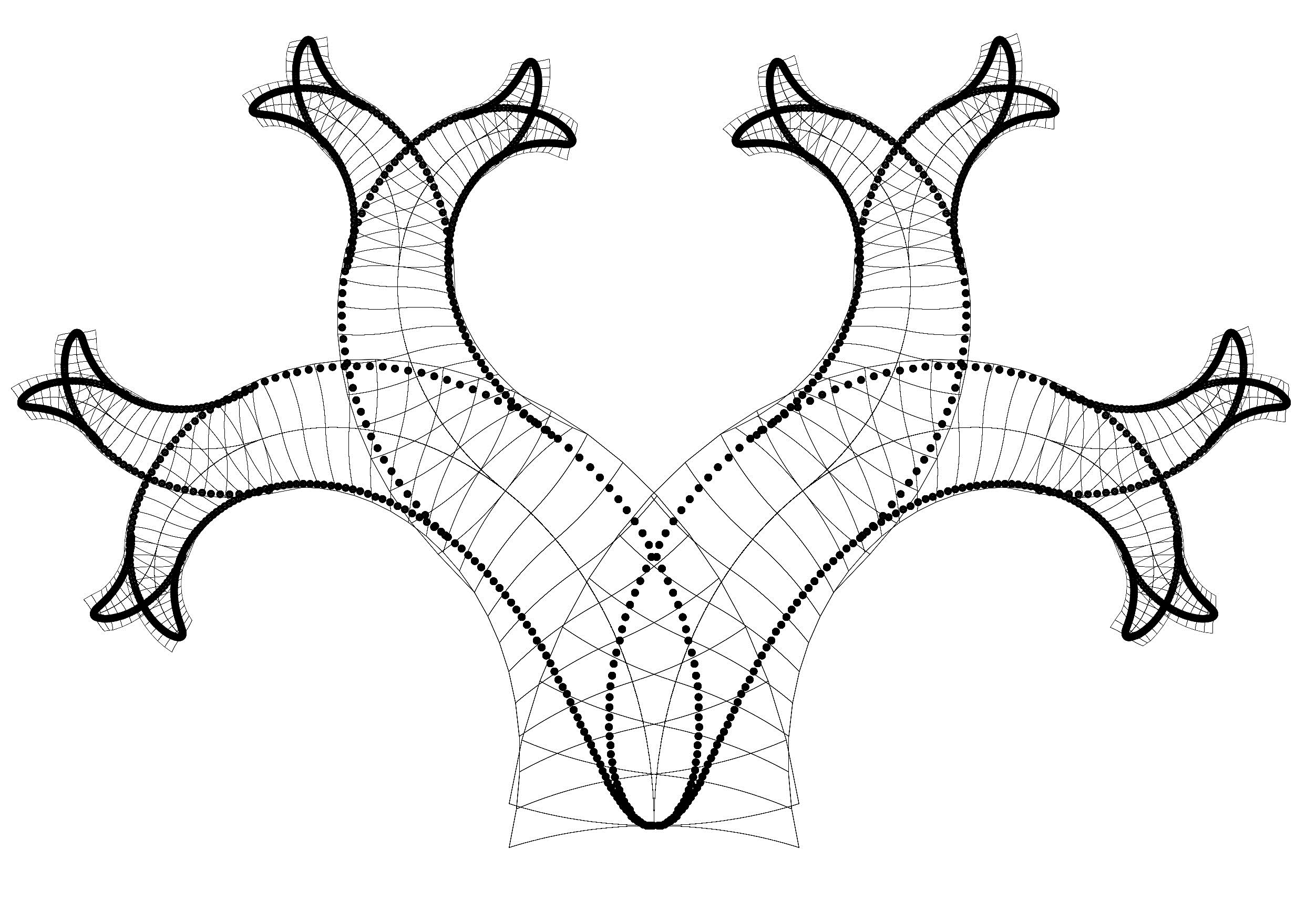}
  \caption{Ellipse plotted on the geometry of an analytic tree fractal}
\end{figure}

Tree fractals may turn and twist around their own branches and the geometry will follow, as shown Image 4 which replicates the first few iterations of the Koch curve. If anything this shows the complexity of basic shapes in a fractal geometric space.

\begin{figure}[!ht]
  \includegraphics[width=1.0\textwidth]{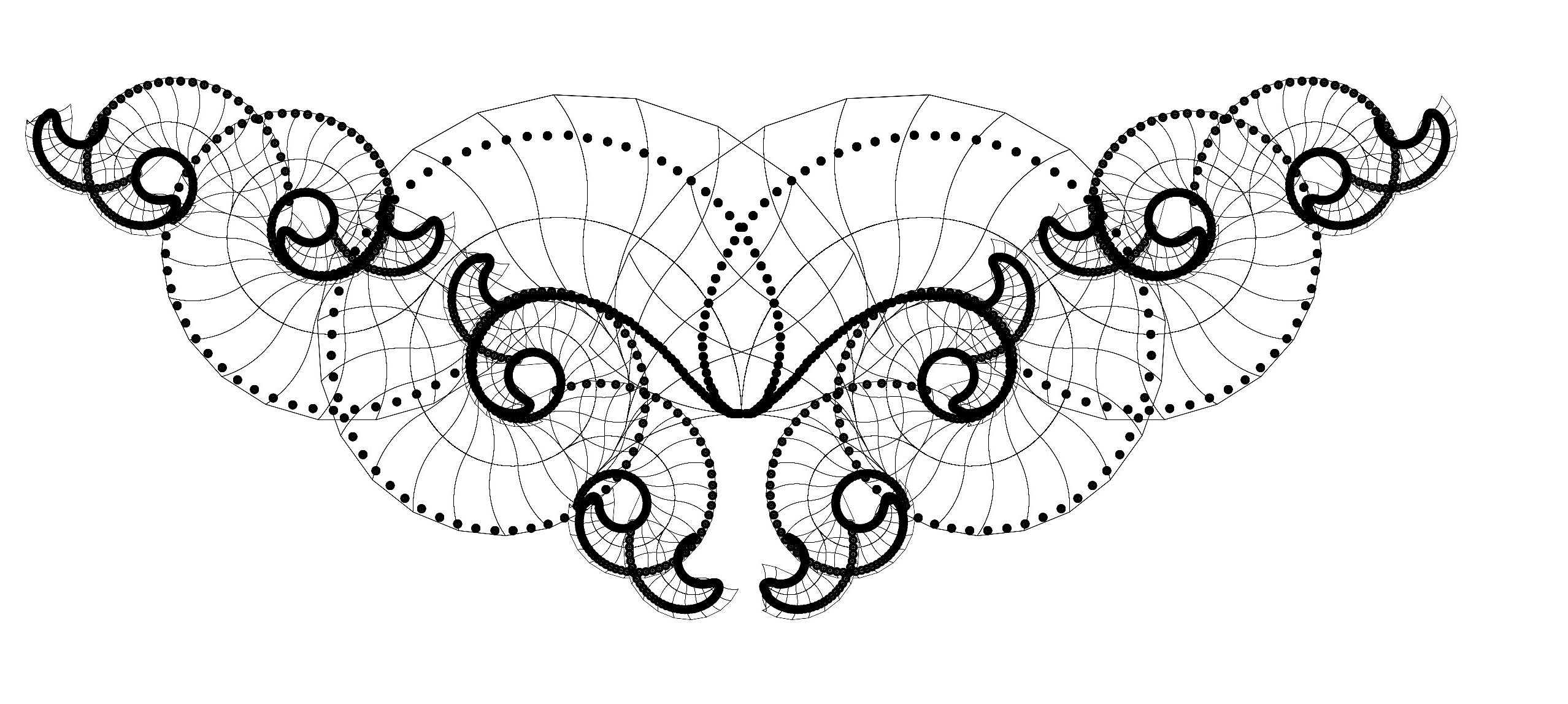}
  \caption{Ellipse plotted on fractal geometry of a Koch curve}
\end{figure}

\section{Canonical form of fractal functions}
One objective in defining a fractal geometry is to study fractal functions in their canonical form, i.e. when the fractal does not branch or scale. This corresponds to the case when $\dot{r}(s)=1$ and $\dot{\varphi}(s)=0$. It would be more accurate to describe this as a fractal where all branches degenerate into overlapping branches. Figure 5 shows the same function as in figure 3 and 4  but with the degenerate condition that  $R=1$ and $\phi=0$. As such we see the function in its canonical form, in this case a parametric ellipse.
\begin{figure}[!ht]
\centering
  \includegraphics[width=0.17\textwidth]{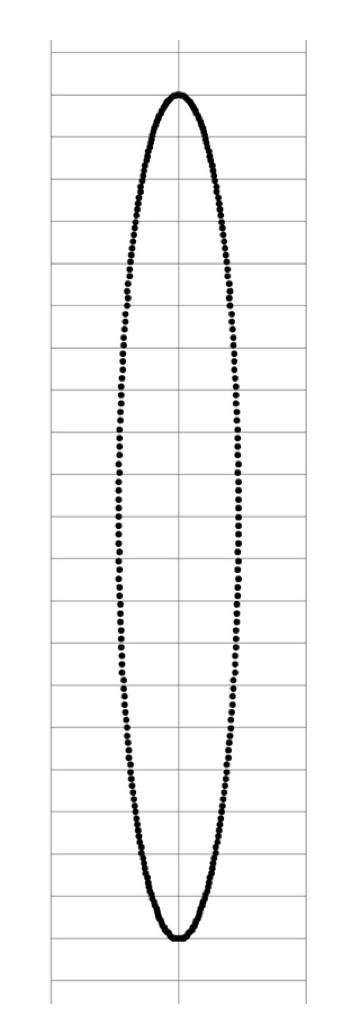}
  \caption{Example of fractal function in its canonical form when $\dot{r}(s)=1$ and $\dot{\varphi}(s)=0$.}
\end{figure}

\section{Adding range restrictions}
From the introduction, we see that there is an overlap between the spaces of adjacent branches. Depending on the values of $\dot{r}(s)$ and $\dot{\varphi}(s)$ the branches may even cross over as they loop around.

In practical applications the overlap of branch space is not always desirable. We can remove the overlap by restricting the range of the fractal function after a branch point. How the range is restricted is arbitrary and the overlap is only avoided locally but this may be sufficient from an application perspective.

Figure 6 shows an example where the range of two symmetric branches has been limited to avoid overlap.

\begin{figure}[!ht]
\centering
  \includegraphics[width=0.3\textwidth]{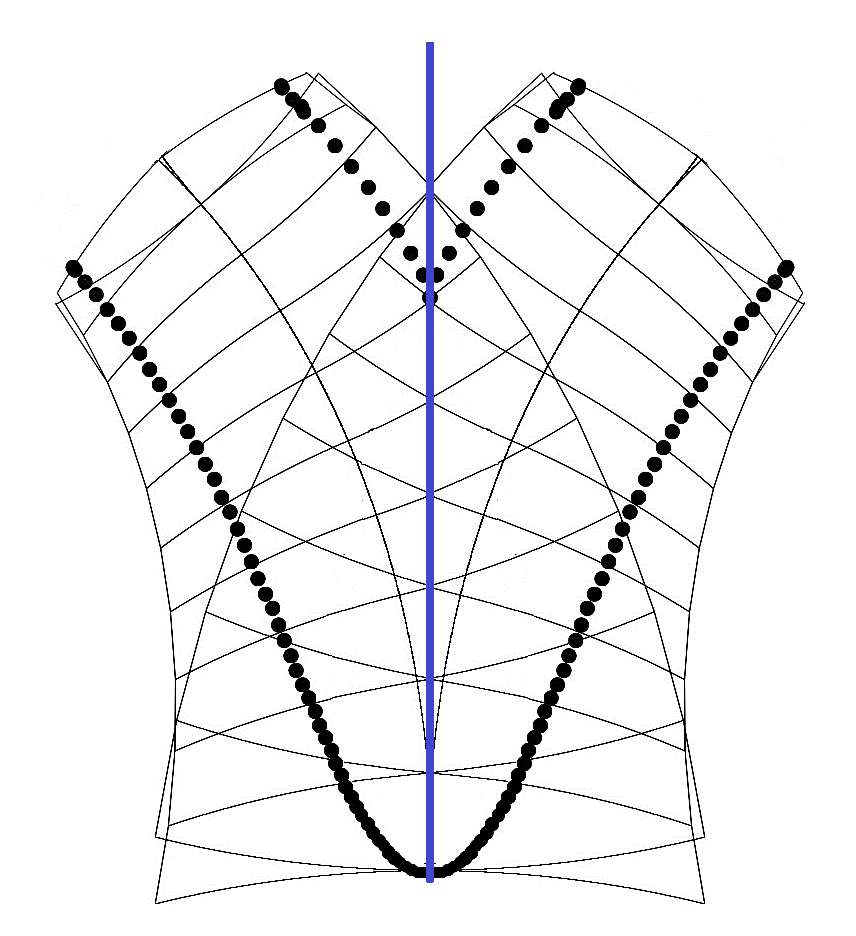}
  \caption{Range restriction (vertical line) between right and left branch to avoid overlapping of the symmetric ellipse branches}
\end{figure}

\section{Extending the fractal geometric space to $\mathbb{R}^n$}
This geometry may be extended to higher dimensions  by using the formulation of derivative coordinate functions as described in [1]. 

The derivative coordinate function in $\mathbb{R}^3$ is given by

\begin{equation}
\check{p}(s)= \left \{ \begin{matrix}\check{p_1}(s)\\ \check{p_2}(s)\\ \check{p_3}(s)\end{matrix} \right\}=\left \{ \begin{matrix}\int \dot{r}\cos(\check{u} \int  \dot{\varphi}ds)\sin(\check{u} \int \dot{\vartheta}ds)ds\\ \int \dot{r}\sin(\check{u}\int \dot{\varphi}ds)\sin(\check{u}  \int \dot{\vartheta}ds)ds\\ \int  \dot{r}\cos(\check{u} \int \dot{\vartheta}ds)ds\end{matrix} \right \}
\end{equation}

where $\check{p_1}(s)$, $\check{p_2}(s)$ and $\check{p_3}(s)$ are the multi-valued Cartesian fractal coordinate functions and $\dot{r}$, $\dot{\varphi}$ and $\dot{\vartheta}$ are the derivative spherical coordinate functions of the fractal tree.

Following a similar argument that led to definition 2.2 and subsequently to equation 7, we construct a path to a general fractal point $\check{f}(x)$

\begin{equation}
\check{f}(x)=\int_{0}^{x_{1}}\check{p}(s)ds + \int_{x_{1}}^{x_{1}+x_{2}}\check{p}(s)ds\cdot \frac{\pi}{2}(\varphi) + \int_{x_{1}+x_{2}}^{x_{1}+x_{2}+x_{3}}\check{p}(s)ds\cdot \frac{\pi}{2}(\vartheta)
\end{equation}

where $\frac{\pi}{2}(\varphi)$ and $\frac{\pi}{2}(\vartheta)$ are $\frac{\pi}{2}$ rotations in the $\varphi$ and $\vartheta$ directions respectively.

This approach may be generalized to any dimension. First we define the derivative coordinate function of a fractal tree $\check{p}$ in  ${\mathbb{R}}^n$ through a recurrence relation with respect to a fractal tree $\check{q}$ in  ${\mathbb{R}}^{n-1}$

\begin{equation}
 \check{p}(s)=\begin{bmatrix}\check{p}_1\\ \vdots \\ \check{p}_{n-1}\\ \check{p}_n\end{bmatrix}=\begin{bmatrix}\check{q}(s)sin(i \check{u}\int \dot{\alpha}_nds)\\ cos(i \check{u}\int \dot{\alpha}_nds)\end{bmatrix}
\end{equation}

where
\begin{equation}
\check{q}(s)=\begin{bmatrix}\check{q}_1\\ \vdots \\ \check{q}_{n-1}\end{bmatrix}
\end{equation}

Now that we have a definition for a fractal tree $\check{p}$ in  ${\mathbb{R}}^n$ we can define a path to a fractal point $\check{x}$ in ${\mathbb{R}}^n$ as follows

\begin{equation}
\check{x}=\begin{bmatrix}\check{x}_1\\ \vdots \\ \check{x}_n\end{bmatrix}=\int_{0}^{x_1} \check{p}(s)ds +\sum_{k=2}^{n}  \left ( \int_{\sum_{x_1}^{x_{k-1}}}^{\sum_{x_1}^{x_k}}\check{p}(s)ds   \cdot \frac{\pi}{2}(\alpha_k) \right )
\end{equation}
where $\frac{\pi}{2}(\alpha_n)$ are $\frac{\pi}{2}$ rotations in the $\alpha_n$  directions.

Now that we have a description of the fractal point $\check{x}$ in ${\mathbb{R}}^n$ by extension we have defined the full fractal geometric space around the smooth tree fractal $\check{p}(s)$ in ${\mathbb{R}}^n$. 
\begin{definition}
The fractal geometric space $\check{F}(x)$ centered around a fractal $\check{p}(x)$ for all coordinates $x=(x_{1},...,x_{n}) \in \mathbb{R}^{n}$ is defined as
\begin{equation}
\check{F}(x)=\int_{0}^{x_1} \check{p}(s)ds +\sum_{k=2}^{n}  \left ( \int_{\sum_{x_1}^{x_{k-1}}}^{\sum_{x_1}^{x_k}}\check{p}(s)ds   \cdot \frac{\pi}{2}(\alpha_k) \right )
\end{equation}
\end{definition}

\section{Uses for this fractal geometry}
As a geometry, this fractal geometry doesn't have the intrinsically useful properties that we find in other geometries. Except in the degenerate case, it is not conformal, as attested by the overlapping geometry of different branches.

Its main usefulness is its ability to allow us to project functions and shapes onto a smooth and continuous tree fractal space. Since such fractals are very common in nature, this may provide a useful tool for studying the fundamental geometry of natural fractals. In particular it allows us to ask questions like: what would a natural form look like in its canonical form when $\dot{r}(s)=1$ and $\dot{\varphi}(s)=0$. 

Without providing evidence, any natural tree would be reduced to a perfectly cylindrical log and leaves would be reduced to discs. River deltas would be represented as canal segments with constant water flow. That last example lifts the veil perhaps. Although care needs to be taken with the transfer of quantities like flow across branch points, the entire fractal system is defined and described by the canonical form i.e. the idealized trunk, the branch "transfer functions" and the fractal coordinate functions $\dot{r}(s)$ and $\dot{\varphi}(s)$ and the fractal unit function $\check{u}(s)$.

To fully make this case, we need to formalize the notion of branch node transfer functions and go beyond symmetric binary fractal trees, i.e. asymmetric n-ary fractal trees. Both of these have been studied and will be the object of separate papers.

\end{document}